\documentclass[twocolumn,prl]{revtex4}
\usepackage{float}
\usepackage{amsfonts}
\usepackage{makeidx}
\usepackage{graphicx}
\usepackage{amsmath,amssymb,graphics,epsfig,color,times,bbm}
\usepackage{CJK}

\begin{document}
\begin{CJK*}{GB}{gbsn}

\title{Graphical rule of transforming continuous-variable graph states by local homodyne detection }
\author{Jing Zhang$^{\dagger }$(Õž¸)}
\affiliation{State Key Laboratory of Quantum Optics and Quantum
Optics Devices, Institute of Opto-Electronics, Shanxi University,
Taiyuan 030006, P.R.China}

\begin{abstract}
Graphical rule, describing that any single-mode homodyne detection
turns a given continuous-variable (CV) graph state into a new one,
is presented. Employing two simple graphical rules: local complement
operation and vertex deletion (single quadrature-amplitude $\hat{x}$
measurement), the graphical rule for any single-mode quadrature
component measurement can be obtained. The shape of CV weighted
graph state may be designed and constructed easily from a given
larger graph state by applying this graphical rule.
\end{abstract}

\maketitle
\end{CJK*}

Graph states - or equivalently called stabilizer states, are a
particularly interesting class of multipartite entangled states
associated with graphs \cite{one,two} and play several fundamental
roles in quantum information such as one-way quantum computation,
quantum error correction, multiparty quantum communication, quantum
simulation, Bell inequalities theorem, etc.

Continuous variables (CV) are a promising new flavor of quantum
information, whose potential is still largely unexplored. CV cluster
and graph states have been proposed \cite{seven}, which can be
generated by squeezed state and linear optics \cite{seven,eight} or
optical frequency comb in an optical parametric oscillator
\cite{nine}, and demonstrated experimentally for the four-mode
cluster state \cite{ten,eleven}. The one-way CV quantum computation
was also proposed with the CV cluster state
\cite{twelve,twelve1,twelve2} and demonstrated with Gaussian
operations experimentally \cite{thirteen}.

It is well known that Clifford operations or projective measurement
associated with operators in the Pauli group can be described by the
stabilizer formalism. For example, the action of local Clifford
operations on qubit \cite{forteen,one} or CV
\cite{fifteen,sixteen,sixteen1} graph states can entirely be
understood in terms of a single elementary graph transformation
rule, called the local complement rule. Moreover, any measurement of
operators in Pauli group turns a given qubit graph state into a new
one \cite{one,two}. The graphical rule also can describe the effect
of Pauli measurement on the qubit graph states. Since CV graph
states have many different properties from the qubit, many concepts
and methods for qubit can not be extended to CV directly. For
example, the CV weighted graph states can be expressed by the
stabilizer formalism \cite{sixteen}. It is distinctively different
from the qubit weighted graph states, which can not be expressed by
the stabilizer formalism. In this paper, how to graphically describe
the effect of local Pauli measurements (local homodyne detection) on
the ideal (i.e., infinitely squeezed) CV weighted graph states is
investigated.

In order to conveniently describe our results, we briefly review
some basic elements of the theory of CV graph states and operations
which are needed in this paper
\cite{seventeen,seven,fifteen,sixteen}. The single mode Pauli
operators are defined as $X(s)=exp[-is\hat{p}]$ and
$Z(t)=exp[it\hat{x}]$, where $s,t\in \mathbb{R}$,
$\hat{x}=(\hat{a}+\hat{a}^\dagger)/\sqrt{2}$ (quadrature-amplitude
or position) and $\hat{p}=-i(\hat{a}-\hat{a}^\dagger)/\sqrt{2}$
(quadrature-phase or momentum). The Pauli operator $X(s)$ is a
position-translation operator, which acts on the computational basis
of position eigenstates as $X(s)|q\rangle=|q+s\rangle$, whereas $Z$
is a momentum-translation operator, which acts on the momentum
eigenstates as $Z(t)|p\rangle=|p+t\rangle$. The Pauli operators for
one mode can be used to construct a set of Pauli operators
$\{X_{i}(s_{i}),Z_{i}(t_{i}); i=1,...,n\}$ for n-mode systems. This
set generates the Pauli group $\mathcal{C}_{1} $. For $k\geq2$, we
can recursively define $\mathcal{C}_{k} $ as \cite{eighteen}
\begin{eqnarray}
\mathcal{C}_{k}=\{U|U\mathcal{C}_{1}U^{\dagger}\subseteq\mathcal{C}_{k-1}\}.
\label{group}
\end{eqnarray}
For every $k$, $\mathcal{C}_{k}\supset\mathcal{C}_{k-1} $, and the
set difference $\mathcal{C}_{k}/\mathcal{C}_{k-1} $ is nonempty.

$\mathcal{C}_{2} $ is a group called the Clifford group, which is
the normalizer of the Pauli group, whose transformations acting by
conjugating, preserve the Pauli group. The Clifford group
$\mathcal{C}_{2} $ for CV is shown \cite{eighteen} to be the
(semidirect) product of the Pauli group and linear symplectic group
of all one-mode and two-mode squeezing transformations.
Transformation between the position and momentum basis is given by
the Fourier transform operator
$F=exp[i(\pi/4)(\hat{x}^{2}+\hat{p}^{2})]$, with
$F|q\rangle_{x}=|q\rangle_{p}$. This is the generalization of the
Hadamard gate for qubits. The phase gate $
P(\eta)=exp[i(\eta/2)\hat{x}^{2}]$ with $\eta\in \mathbb{R}$ is a
squeezing operation for CV and the action $P(\eta)RP^{-1}(\eta)$ on
the Pauli operators is
\begin{eqnarray}
P(\eta):X(s)&\rightarrow& e^{-is^{2}\eta/2}Z(s\eta)X(s)
,\nonumber \\
Z(t)&\rightarrow& Z(t),  \label{phase}
\end{eqnarray}
in analogy to the phase gate of qubit. The controlled operation C-Z
is generalized to controlled-$ Z (C_{Z})$. This gate
$C_{Z}=exp[i\hat{x}_{1}\bigotimes\hat{x}_{2}]$ provides the basic
interaction for two mode 1 and 2, and describes the quantum
nondemolition (QND) interaction. This set $ \{X(s), F, P(\eta), C-Z;
s,\eta \in \mathbb{R}\}$ generates the Clifford group. Here the
controlled operation with any interaction strength
$C_{Z}(\Omega)=exp[i\Omega\hat{x}_{1}\bigotimes\hat{x}_{2}]$
($\Omega\in \mathbb{R}$) will be used in this paper. Another type of
the phase gate will also be utilized $
P_{X}(\eta)=FP(\eta)F^{-1}=exp[i(\eta/2)\hat{p}^{2}]$ and the action
on the Pauli operators is
\begin{eqnarray}
P_{X}(\eta):X(s)&\rightarrow& X(s)
,\nonumber \\
Z(t)&\rightarrow& e^{-it^{2}\eta/2}X(-t\eta)Z(t),  \label{phase1}
\end{eqnarray}
where $P_{X}(\eta)^{\dagger}=P_{X}(\eta)^{-1}=P_{X}(-\eta)$.

A weighted graph quantum state is described by a mathematical graph
$G=(V,E)$, i.e. a finite set of $n$ vertices $V$ connected by a set
of edges $E$ \cite{forteen}, in which every edge is specified by a
factor $\Omega_{ab}$ corresponding to the strength the modes a and
b. The preparation procedure of CV weighted graph states is only to
use the Clifford operations: first, prepare each mode (or graph
vertex) in a phase-squeezed state, approximating a zero-phase
eigenstate (analog of Pauli-X eigenstates), then, apply the QND
coupling ($C_{Z}(\Omega)$) with the different interaction strength
$\Omega_{jk}$ to each pair of modes $(j,k)$ linked by a weighted
edge in the graph. The CV unweighted graph states is to use the QND
interaction all with the same strength. Since all C-Z gates commute,
the resulting CV graph state becomes, in the limit of infinite
squeezing, $g_{a}=(\hat{p}_{a}-\sum_{b\in
N_{a}}\Omega_{ab}\hat{x}_{b})\rightarrow0$, where the modes $a\in V$
correspond to the vertices of the graph of $n$ modes, while the
modes $b\in N_{a}$ are the nearest neighbors of mode $a$. This
relation is as a simultaneous zero-eigenstate of the
position-momentum linear combination operators. The corresponding
$n$ independent stabilizers for CV weighted graph states are
expressed by $G_{a}(\xi)=exp[-i \xi g_{a}]=X_{a}(\xi)\prod_{b\in
N_{a}}Z_{b}(\Omega_{ab}\xi)$ with $\xi\in \mathbb{R}$.

The local complement operation on vertex $a$ of CV weighted graph
state is expressed by  \cite{sixteen}
\begin{eqnarray}
U_{LG_{a}}(\delta)=P_{Xa}(-\delta)\prod_{b\in
N_{a}}P_{b}(\Omega_{ab}^{2}\delta). \label{LC}
\end{eqnarray}
The action of this local complement operation on CV weighted graph
states can be stated in purely graph theoretical terms, which
completely characterize the evolution of CV weighted graph states
under this local complement operation. The graph rule of applying
this local complement operation on vertex $a$ is described as: first
obtain the subgraph of $G$ generated by the neighborhood $N_{a}$ of
$a$, then reset the weight factor of all edges of this subgraph
calculated with the equation
$\Omega_{b_{i}b_{j}}'=\Omega_{b_{i}b_{j}}-\Omega_{ab_{i}}\Omega_{ab_{j}}\delta$,
at last delete all the edges with the weight factor of zero, and
leave the rest of the graph unchanged. Here, a subgraph $G[C]$ of a
graph $G=(V,E)$, where $C\subset V$, is obtained by deleting all
vertices and the incident edges that are not contained in $C$.

First, we consider the simplest case of the quadrature-amplitude
($\hat{x}$) component measurement on vertex $a$ of CV weighted graph
state. After the $\hat{x}$ component measurement on vertex $a$ with
the result $x_{a}$ by local homodyne detection, a CV weighted graph
state $|G\rangle$ will transform into another weighted graph state
$|G'\rangle=U_{x_{a}}^{(a)}|G-a\rangle$, where $G-a$ denote the
graph that is obtained from $G$ by deleting the vertex $a$ and all
edges incident with $a$, and
\begin{eqnarray}
U_{x_{a}}^{(a)}=\prod_{b\in N_{a}}Z^{(b)}(x_{a}). \label{trans}
\end{eqnarray}
Thus, in case of a measurement of quadrature-amplitude ($\hat{x}$),
the resulting graph can be produced by simply deleting the measured
vertex $a$ and related edges from the graph, then translating the
measurement result into the momentum component of the neighborhood
vertices $N_{a}$ of $a$. Based on above two simple graphical rules:
local complement operation and the single quadrature-amplitude
$\hat{x}$ measurement (also called vertex deletion), the graphical
rules for any quadrature component measurement can be obtained as
the following.

\begin{figure}

\centerline{
\includegraphics[width=3in]{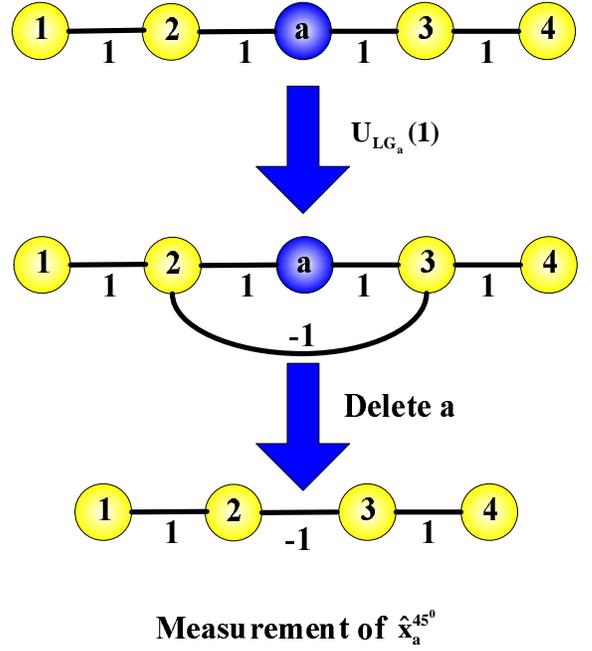}
} \vspace{0.1in}

\caption{\label{Fig:epsart} Example of a $\hat{x}^{45^{0}}$
measurement on vertex a in a linear cluster.  }
\end{figure}

Any quadrature component of an optical mode is expressed by
\begin{eqnarray}
\hat{x}^{\theta }&=&(\hat{a}e^{-i\theta
}+\hat{a}^{\dagger}e^{i\theta
})/\sqrt{2}\notag \\
&=& \cos\theta\hat{x}+\sin\theta\hat{p}. \label{qua}
\end{eqnarray}
The basic idea of obtaining the graphical rules for any quadrature
component measurement is that, first transform the
quadrature-amplitude $\hat{x}$ of vertex $a$ into the measured
component $\hat{x}'=\cos\theta\hat{x}+\sin\theta\hat{p}$ by local
Gaussian operations, then perform single quadrature-amplitude
$\hat{x}$ measurement (vertex deletion). Now we first apply local
complement operation $U_{LG_{a}}(\delta)$ on vertex $a$ of CV
weighted graph state. Thus the quadrature-amplitude of vertex $a$ is
transformed into $\hat{x}_{a}'=\hat{x}_{a}+\delta\hat{p}_{a}$, where
$\delta=\tan \theta$. Then we perform the quadrature-amplitude
measurement on vertex $a$ with the detection result $x_{a}'$. The
new graph is obtained by deleting the measured vertex $a$ and
related edges from the graph, then translating the measurement
result $x_{a}'$ into the momentum component of the neighborhood
vertices $N_{a}$ of $a$. The new graph can be expressed by
$|G''\rangle=U_{x'_{a}}^{(a)}|G'-a\rangle$ and
$|G'\rangle=U_{LG_{a}}(\tan\theta)|G\rangle$. Figure 1 presents an
example of a $\hat{x}^{45^{0}}$ measurement on a linear cluster,
which removes the measured mode and still keeps the neighbor
vertices to be joined.

\begin{figure}

\centerline{
\includegraphics[width=3in]{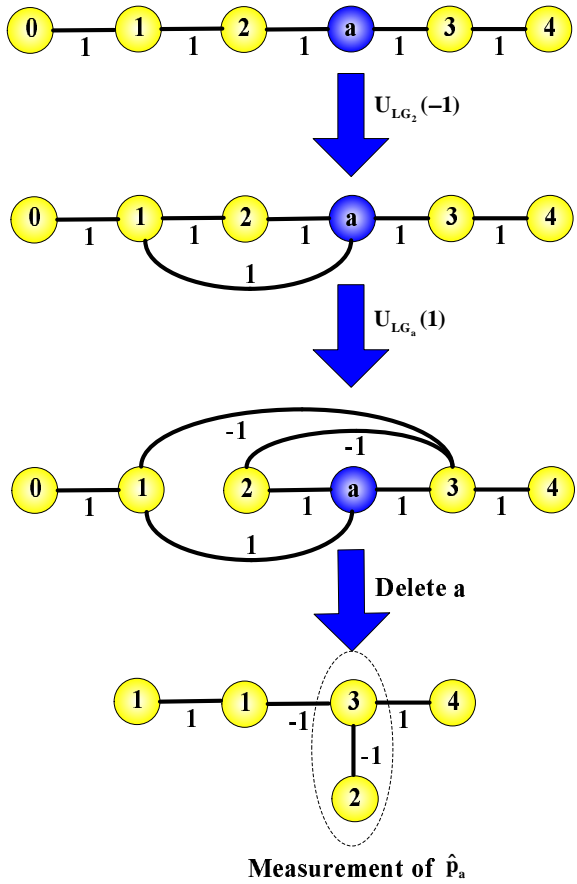}
} \vspace{0.1in}

\caption{\label{Fig:epsart} Example of a $\hat{p}$ measurement on
vertex a on a linear cluster. The vertex 2 was chosen as the special
neighbor $b_{0}$. }
\end{figure}

Now considering the single quadrature-phase $\hat{p}$ measurement,
it requires that $\delta$ approach infinite for local complement
operation $U_{LG_{a}}(\delta)$ on vertex $a$. Apparently it is
impractical. So we consider the different local operations to obtain
the graphical rule for quadrature-phase component measurement.
First, chooses any of neighbor vertex $b$ of vertex $a$, and applies
local complement operation $U_{LG_{b}}(-1/\Omega_{ab}^{2})$ on
vertex $b$. Then applies local local complement operation
$U_{LG_{a}}(1)$ on vertex $a$. Thus the quadrature-amplitude and
phase of vertex $a$ are transformed into $\hat{x}_{a}'=\hat{p}_{a}$,
$\hat{p}_{a}'=\hat{p}_{a}-\hat{x}_{a}$ by $P_{Xa}(-1)P_{a}(-1)$. The
new graph can be expressed by
$|G''\rangle=U_{x'_{a}}^{(a)}|G'-a\rangle$ and
$|G'\rangle=U_{LG_{a}}(1)U_{LG_{b}}(-1/\Omega_{ab}^{2})|G\rangle$.
An example of a $\hat{p}$ measurement on a linear cluster is
depicted in Fig. 2. It is easy to see that two adjacent vertices
combine into a single vertex with the bonds attached to each. Figure
3 presents a $\hat{p}$ measurement on a complex weighted graph
state.

\begin{figure}

\centerline{
\includegraphics[width=3in]{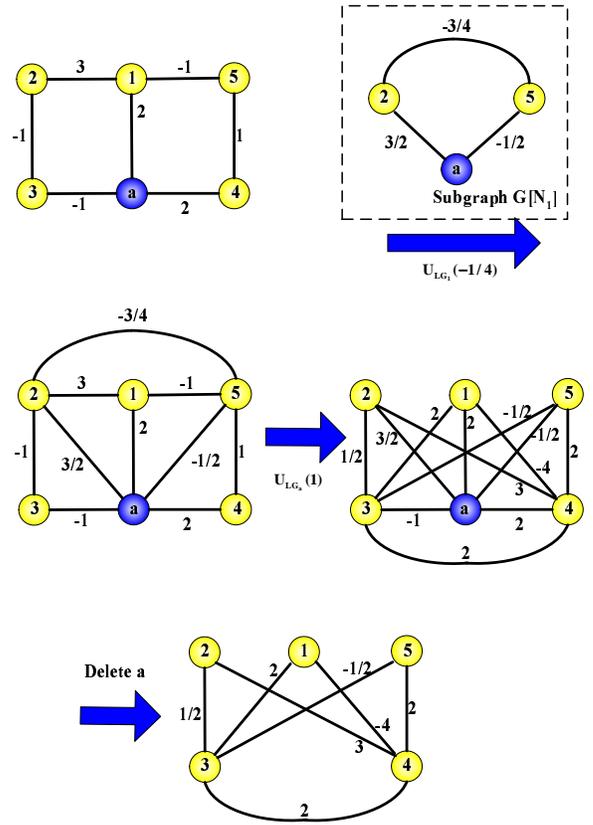}
} \vspace{0.1in}

\caption{\label{Fig:epsart} The graphical rule for a $\hat{p}$
measurement on a complex weighted graph state. The vertex 1 was
chosen as the special neighbor $b_{0}$.}
\end{figure}

In summary, graphical rule of transforming ideal CV weighted graph
state by any single-mode homodyne detection is described. The
graphical rule for any single-mode quadrature component measurement
is obtained by two simple graphic rules: local complement operation
and vertex deletion. This work will be very helpful to specify which
graph states can mapped by the single-mode homodyne detection and to
apply in one-way CV quantum computation.

$^{\dagger} $Corresponding author's email address:
jzhang74@sxu.edu.cn, jzhang74@yahoo.com

\section{\textbf{ACKNOWLEDGMENTS}}

This research was supported in part by NSFC for Distinguished Young
Scholars (Grant No. 10725416), National Basic Research Program of
China (Grant No. 2006CB921101, 2011CB921601), NSFC Project for
Excellent Research Team (Grant No. 60821004), and Program for the
Top Young and Middle-aged Innovative Talents of Higher Learning
Institutions of Shanxi.

$Note$ $added.$ - Right after completion of this work, another work
appeared \cite{ninteen}, where two simple and special local homodyne
measurements were investigated experimentally.

\end{document}